# Exploiting the Benefits of V2B Application on Peak Shaving of Data Center Loads

Arya Joshi
Power and Energy Division
WSP USA
Irvine, CA, USA
arya.joshi@wsp.com

Hamed Haggi
Power and Energy Division
WSP USA
Houston, TX, USA
hamed.haggi@wsp.com

Chinmay Morankar
Power and Energy Division
WSP USA
Houston, TX, USA
chinmay.morankar@wsp.com

***Abstract*****—**The accelerated growth in data center projects has introduced a demand-driven bottleneck throughout power grids and contributed to a substantial increase in carbon emissions. These concerns are fueling discussions on methods to use existing energy assets to drive operational efficiency. To this end, this paper explores the benefits of Vehicle-to-Building (V2B) applications to support peak shaving of data center cooling loads. Initially, a literature review was conducted considering V2B constraints and optimization methods including SoC limitations, EV participation, tariffs, and building loads. This analysis was then used to develop a conceptual case study of a 10 MW data center in Loudoun County, VA by simulating a temperature-dependent load profile and adjusting the V2B participation of 40 commercial and passenger EVs. Simulation results indicate that, depending on seasonal variations in cooling load demands, strategic deployment of V2B assets between 12-5pm can offset gross cooling loads by 13-36%.

***Keywords— Electric Vehicles, Data Center, Microgrid, Peak Shaving, Vehicle-to-Building.***

## I. Introduction

As computational demand has increased with the advent of artificial intelligence, hyperscalers have been investing heavily in data centers. In fiscal 2024, commitments to AI infrastructure spending by just a few US hyperscalers (ex. Microsoft, Amazon, etc.) amounted to 320 billion dollars [1]. This massive surge in demand is affecting power grids in unprecedented ways. In Loudoun County, VA, often described as the "data center capital of the world," utilities are slowing or canceling the decommissioning of fossil fuel generation plants just to support the 200 data centers currently operational and 117 in the development pipeline [2]. Overall, this expansion is leading to an increase in electricity demand, volatility in cleared market price, reliability and resilience issues, and carbon emissions. For instance, training the OpenAI GPT-3 LLM model consumed over 1.287 GWh of energy and produced 552 tons of carbon emissions [3]. To address the increase in demand and the adverse impact on the environment, it is pertinent to explore methods of managing energy that can address peak loads, increase renewable energy penetration, and reduce grid loads [4].

Vehicle-to-building (V2B) infrastructure is a promising candidate for adding operational and energy efficiencies to data centers without the need for installing and deploying expensive battery energy storage system (BESS) assets. Electric vehicles (EVs) can provide controlled power delivery to commercial buildings and/or operations using bi-directional chargers to offset peak loads. Its adoption, along with a behind-the-meter energy management system (EMS), can improve grid reliability and renewable energy penetration. Moreover, using V2B for peak load shaving not only reduces grid reliance, but also reduces peak charge costs and demand charges [5]. In data center applications, V2B can specifically help shave peak cooling loads, allowing data centers to be more energy efficient, resilient, and cost-effective. However, the feasibility and techno-economic operational constraints of V2B integration in the data center use case have largely been unaddressed. Motivated by this challenge, this paper proposes a framework that exploits the benefits of V2B applications for peak shaving of data center loads. In addition, real world simulation and techno economic operational analysis have been done to support the framework.

The rest of the paper is organized as follows: Section II will discuss the proposed V2B framework for data center application and related literature review, Section III will focus on simulation results and numerical analysis, and finally section IV concludes the paper and presents some future directions of this study.

## II. V2B Framework and Literature Review

EV participation and state-of-charge (SoC) constraints are foundational in V2B studies, as they determine dispatchable energy capacity. Yoo et al. [6] modeled five EVs with variable SoC supporting a 70-kW data center, achieving a 10% energy cost reduction, though without accounting for dynamic EV participation. Since temporal EV availability strongly influences usable capacity, Ahsan et al. [7] employed a Monte Carlo–based arrival model and reported more realistic daily savings of 4–9%. These participation levels also drive infrastructure investment needs. Using MILP, He et al. [5] estimated that supporting a 30-kW peak load requires 8–29 EVs in summer and 2–15 in winter, reflecting higher summer cooling demand. Effective V2B dispatch further depends on predictive, weather-sensitive load modeling, as shown by Ghafoori et al. [8], who emphasized the role of HVAC and geographic conditions. Technoeconomic assessments must also consider financial incentives: Osman et al. [9] demonstrated that tariff-based peak shaving and availability incentives can offset battery degradation costs, while Luo et al. [10] showed V2B can increase renewable penetration by 45.1%, reduce LCOE by up to 38.9%, and reduce net energy costs for EV owners by 18.1%.

Collectively, these studies show that V2B can reduce peak loads, electricity costs, LCOE, and EV ownership costs, provided feasibility and optimization analyses incorporate SoC constraints, EV participation, fleet size, and temperature-dependent load profiles.

## III. Simulation Results and Numerical Analysis

Integrating V2B architecture into data center operations can enable peak cooling load shaving. This use case was evaluated through a case study of a 10 MW data center in Leesburg, VA (Loudoun County), a region with rapid data center growth. Cooling demand was modeled using a time-varying PUE between 1.2 and 1.4, resulting in cooling loads ranging from 2 MW ($P_{min}$) at 4 a.m. to 4 MW ($P_{max}$) at 4 p.m. The baseline cooling load ($P_b$) was represented using a cosine function to capture diurnal variation (1).

$$P_b = \frac{(P_{max}+P_{min})}{2} + \frac{(P_{max}-P_{min})}{2} \cdot \cos\left(\frac{\pi}{12} \cdot (t-16)\right) \tag{1}$$

The base temperature was assumed to be 20 C ($T_b$). However, to incorporate the impact of temperature on cooling loads, an 8760-temperature profile ($T_r$) for Loudoun County was acquired from the NREL National Solar Radiation Database [11]. A cooling variation value ($\alpha$) of 0.03 kW/C was applied to create a temperature dependent hourly load profile ($P_t$) (2).

$$P_t = P_b \cdot (1+\alpha) \cdot (T_r - T_b) \tag{2}$$

Additionally, it was assumed that the data center had 20 company-owned Ford F-150 Lightning trucks (Table 1). While these were assumed to be unavailable for V2B participation during early operating hours, they would be available to actively participate after 3 pm with an average SoC of 80%. The EVs would continue discharging at the specified power rating until all available capacity is fully utilized. It was also assumed that the total number of participating on-site passenger EVs ($V_p$), modeled using Tesla Model 3 parameters (Table 1), followed a cosine curve strictly from 8 am to 6 pm, with a peak of 20 vehicles ($V_{max}$) at 1 pm (3). This attempted to model real-world extended operating hours of data centers. Moreover, it was presumed that the number of participating passengers in EVs had full energy capacity that could be deployed at any given time.

$$\frac{V_{max}}{2}\left(1+\cos\left(\frac{\pi}{5}(t-13)\right)\right),\ 8\le t\le 18,\ 0\ otherwise \tag{3}$$

TABLE I - VEHICLE INPUT PARAMETERS

| Vehicle | Vehicle Properties | | | | |
|---|---|---|---|---|---|
| | ***Energy Capacity (kWh)*** | ***Power Rating (kW)*** | ***C-Rate*** | ***Participation*** | ***Max. number of vehicles*** |
| Ford F-150 Lightning [12] | 131 | 17.6 | 0.13 | Full | 20 |
| Tesla Model 3 [13] | 80 | 24 | 0.3 | Sinusoidal | 20 |

Using the temporal, temperature-dependent cooling load profile, EV mix and participation schedule, and the SoC constraints, the hourly net load profile was calculated for an average winter day (1/1) and average summer day (8/1) in the TMY (Figure 1 (a) and (b)).

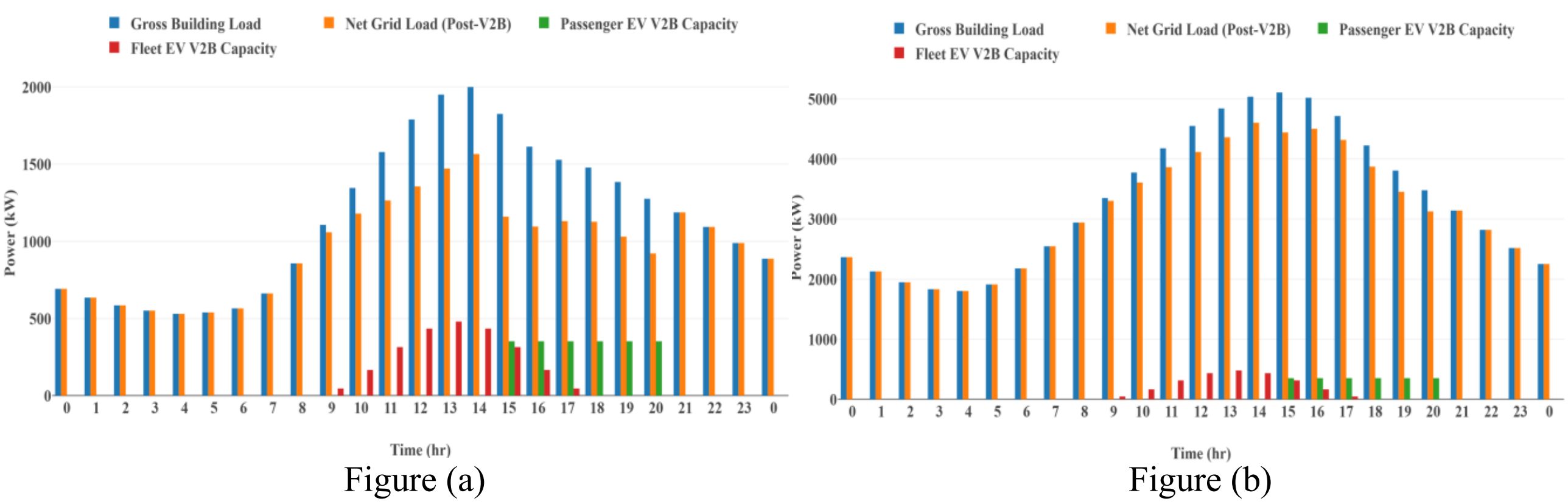

Fig. 1. Peak load shaving of 10 MW data center (a) Winter Day (b) Summer Day

As seen in both figures, V2B does not contribute to the offset of cooling loads until 9 am, when passenger EV participation begins to increase. The greatest difference between net and gross cooling load occurs in the afternoon between 12-5 pm as a result of peak passenger and fleet EV availability. This trend is expected as company-owned EVs are usually available after completing routine maintenance work in early morning/afternoon, and most employees owning passenger EVs would already be on-site. Moreover, this participation schedule aligns well with peak cooling loads which occur during late afternoons when temperatures are highest. The figures indicate that daily gross loads can be reduced by approximately 13-36% by strategically deploying V2B in the afternoon. Load offset is pronounced during the winter season given the overall cooling demand is lower whereas higher cooling demand in summer reduces the relative effect of V2B contributions.

## IV. Conclusions and Future Works

Rapid AI-driven data center growth is straining power grids and increasing emissions, accelerating interest in leveraging existing energy assets such as EVs and batteries. This paper proposes using a corporate EV fleet, Ford F-150 Lightnings and Tesla Model 3s, for V2B deployment to offset data center cooling loads. Simulated summer and winter profiles indicate that strategic V2B dispatch can offset approximately 13–36% of peak cooling demand. Future work will assess how tariff structures and EV incentives affect energy cost savings for data center applications.

## References

[1] I. Hitchcock and M. Cahoon, *Hyperscaler Data Center Buildout: A Sustainability Bane, Boon, or Both?* NI R 25-04. Durham, NC, USA: Nicholas Institute for Energy, Environment & Sustainability, Duke University, 2025. [Online]. Available: https://nicholasinstitute.duke.edu/publications.

[2] Loudoun County, VA, and M. Turner, *Loudoun County, Virginia: Data Center Capital of the World—A Strategy for a Changing Paradigm*, 2024.

[3] Z. Ji, et al., “A systematic review of electricity demand for large language models: Evaluations, challenges, and solutions,” *Renew. and Sustainable Energy Reviews*, vol. 225, p. 116159, 2026.

[4] P. Parikh *et al*., “Strain to synergy: Powering AI data centers with sustainable and smart power grid infrastructures,” in *Proc. IEEE Int. Conf. Commun., Control, and Computing Technologies for Smart Grids (SmartGridComm)*, 2025, pp. 1–6.

[5] Z. He, J. Khazaei, and JD. Freihaut, "Optimal integration of Vehicle to Building (V2B) and Building to Vehicle (B2V) technologies for commercial buildings." *Sustainable Energy, Grids and Networks* 32 (2022): 100921.

[6] Y. Yoo and A. Tchagang, “V2B/V2G on energy cost and battery degradation under different driving scenarios, peak shaving, and frequency regulation,” *World Electric Vehicle Journal*, vol. 11, no. 1, p. 14, 2020.

[7] S. M. Ahsan *et al*., “Optimized power dispatch for smart buildings and electric vehicles with V2V, V2B, and V2G operations,” *Energies*, vol. 16, no. 13, p. 4884, 2023.

[8] M. Ghafoori *et al*., “Electricity peak shaving for commercial buildings using machine learning and vehicle-to-building (V2B) systems,” *Applied Energy*, vol. 340, p. 121052, 2023.

[9] H. J. Osman *et al*., “Vehicle-to-building (V2B) peak load shaving and tariff analysis,” in *IOP Conf. Ser.: Earth Environ. Sci.*, vol. 1395, no. 1, p. 012019, 2024.

[10] J. Luo *et al*., “Comprehensive techno-economic performance assessment of PV–building–EV integrated energy systems concerning V2B impacts on both building energy consumers and EV owners,” *Journal of Building Engineering*, vol. 87, p. 109075, 2024.

[11] National Renewable Energy Laboratory (NREL). n.d. “National Solar Radiation Database (NSRDB) Data Viewer.” Accessed April 28, 2026. https://nsrdb.nlr.gov/data-viewer.

[12] Ford Motor Company, *Ford F-150 Lightning Technical Specifications*. From the Road, 2025.

[13] Tesla, Inc., *Model 3 Owner’s Manual*, 2026.